# Excitonic Emission in Organic-Inorganic Lead Iodide Perovskite Single Crystals.

Hiba Diab[a], Gaëlle Trippé-Allard[a], Ferdinand Lédée[a], Khaoula Jemli [a], Christèle Vilar [b], Guillaume Bouchez[b], Vincent L.R. Jacques[c], Antonio Tejeda[c], Jean-Sébastien Lauret[a], Emmanuelle Deleporte*[a], Damien Garrot*[b]

[a] *Laboratoire Aimé Cotton, Ecole Normale Supérieure de Cachan, CNRS, Université de Paris-Sud, Université Paris-Saclay, Bât 505 Campus d'Orsay, 91405 Orsay, France*

[b] *Groupe d'Etude de la Matière Condensée, Université de Versailles Saint Quentin En Yvelines, Université Paris-Saclay, 45 Avenue des Etats-Unis, 78035, Versailles, France.*

[c] *Laboratoire de Physique des Solides, Université Paris-Sud, Université Paris-Saclay, Bât 510, 91405 Orsay, France*

**Corresponding Authors**

*Email:

damien.garrot@uvsq.fr

emmanuelle.deleporte@ens-cachan.fr

# Abstract

Hybrid perovskite thin films have demonstrated impressive performance for solar energy conversion and optoelectronic applications. However, further progress will benefit from a better knowledge of the intrinsic photophysics of materials. Here, the low temperature emission properties of $CH_3NH_3PbI_3$ single crystals are investigated and compared to those of thin polycrystalline films by means of steady-state and time-resolved photoluminescence spectroscopy. While the emission properties of thin films and crystals appear relatively similar at room temperature, low temperature photoluminescence spectroscopy reveals striking differences between the two materials. Single crystals photoluminescence exhibits a sharp excitonic emission at high energy, with Full Width at Half Maximum of only 5 meV, assigned to free excitonic recombination and a broad band at low energy. We analyzed the thermal evolution of the free excitonic intensity and linewidth. An excitonic binding energy of 28 meV is extracted from the quenching of the photoluminescence. We highlight a strong broadening of the emission due to LO phonons coupling. The free excitonic emission turned to be very short-lived with a sub-nanosecond dynamics, mainly induced by the fast trapping of the excitons. The free excitonic emission is completely absent of the thin films spectra, which are dominated by trap states band. The trap states energies, width and recombination dynamics present important similarities between films and crystals. These results suggest that the trap states are formed at the surface and grain interface of the perovskite.

## 1. Introduction

Recently, a great attention has been devoted to three-dimensional hybrid organic perovskites (HOP) due to their remarkable results as light-harvesting material for low-cost solar cells. In particular, a breakthrough has been achieved with solar energy conversion reaching 22% using hybrid perovskite as active material.[1–3] Moreover, HOP demonstrate impressive emission properties which make them an attractive material for light emitting diodes and laser applications.[4–6]

A better understanding of the photophysical properties underlying these performances is important for further device improvements. The study of the photoluminescence (PL) of semiconductors at low temperature has proved to be an efficient tool not only to understand the material physics but also to assess the crystalline quality, the nature of defects, and impurities. Studies on the emission of $CH_3NH_3PbI_3$ at low temperature are relatively scarce and the majority concerns polycrystalline thin films.[7–10] However, the grain structure of the perovskites thin films have a strong impact on the optical properties: the band gap position, the carrier diffusion, the recombination and the excitonic effects depend on the degree of crystallinity of the sample.[11–14] Extrinsic defects formed at the interface of the grains could play an important role in the optical and electronic properties of thin films.[15,16] Hence, in order to measure the intrinsic characteristics of HOP, the study of bulk single crystals is necessary. The crystal has not the flexibility of solution-processed thin films for the realization of devices, but should serve as a reference material. However, to the best of our knowledge, the emission properties of a single crystal of $CH_3NH_3PbI_3$ at low temperature has only been reported by Fang *et al.*[17]

Among the fundamental properties of HOP, the nature of the photoexcited states, excitons or free charges, is an important question to understand the remarkable performance of $CH_3NH_3PbI_3$ in solar cells and optoelectronic devices and has been discussed in numerous studies.[7,8,18–22] Recent studies suggest that free carriers are the dominant species at room temperature, while excitons prevail in the low temperature orthorhombic phase. Excitons could generally be observed in bulk semiconductors through the emission of sharp lines at low temperature with Full Width at Half Maximum (FWHM) of ca. 1 meV.[23] However, previous studies on both films and single crystals report multi-component, broad emissions, with a FWHM of several tens of meV, presenting a weak broadening with temperature.[8–10,17,24,25]

In this study, we performed temperature-dependent Steady-State and Time-Resolved Photoluminescence on high quality $CH_3NH_3PbI_3$ single crystals. We demonstrate the existence of important excitonic effects on the emission properties at low temperature, compared to the emission of polycrystalline thin films. Single crystal presents at high energy, a narrow emission of only 5 meV at 10K, which is absent of the PL spectra of films. From the position, line shape and power dependence, we attribute this emission to free excitonic recombination. Analysis of the thermal evolution of this emission results in an estimation of the exciton binding energy of ca. 28 meV and demonstrates a strong homogeneous broadening induced by LO phonon coupling. Besides, this emission presents a non-exponential, fast recombination dynamic, with sub-nanosecond leading time. A low energy band is present on both films and crystal and share similar energy positions, and recombination dynamics. This low energy band is connected to trap states, likely formed at the surfaces/interfaces of materials.

## 2. Experimental

**Thin films:** Methylammonium iodide $CH_3NH_3I$ (called MAI hereafter) is synthetized by reacting 20 mL $CH_3NH_2$ (2M in methanol, Sigma Aldrich) and 10,5 mL aqueous solution of iodic acid HI (57% in water, Sigma Aldrich) in a 250mL round –bottom flask stirred at 0°C for 2 hours. The precipitate is recovered after evaporation at 60°C for 1 hour using a rotary evaporator. After several washing in diethyl ether, the white MAI microcrystals were dried overnight at 60°C. To prepare the perovskite $CH_3NH_3PbI_3$, 596 mg (5mmol) of the synthesized MAI and 1,729 g (5mmol) of lead iodide $PbI_2$ (Sigma Aldrich 99%) are mixed in 5 mL of dimethylformamide (DMF, Sigma Aldrich reagent plus >99%) and stirred at 60°c for 12 hours (1:1 molar ratio of precursors). The obtained yellow solution is homogeneous with high transparency in the visible region.

UV quartz slides (from Neyco) are used as substrates. They are cleaned with acetone and 2-propanol in ultrasonic bath for 15 minutes each. The substrates are then cleaned with a solution of 10%w KOH in ethanol for 15 minutes. The slides are then rinsed with distilled water and dried with pressured air. $CH_3NH_3PbI_3$ thin films are obtained by spin-coating the precursor solution in DMF at 2000rpm for 15 seconds. After the deposition, the samples are annealed at 95°C for 20 minutes. The thicknesses of the layers have been determined by profilometry. The film presents the expected morphology for pure $CH_3NH_3PbI_3$ thin films, prepared with one-step fabrication protocols (Figure 1(c)).[26]

**Single crystals:** Single crystals have been synthetized following previously reported procedure.[27–30] Methylammonium iodide (0.78 g, 5 mmol) and lead iodide (2.30 g, 5 mmol) were dissolved in GBL (5 mL) at 60°C. 2 mL of the yellow solution was placed into a vial and heated at 120°C during one to four hours depending on the crystals size expected. The solution could be heated in a hot plate as well as an oil bath. XRD pattern is reported in Supplementary Information.

**Scanning electron microscopy:**

The crystal was placed on conductive carbon tape. SEM views were acquired using a commercial JEOL JSM-7001F field emission SEM.

**X-ray diffraction**: XRD (X-Ray Diffraction) patterns were obtained on a 1x1x1 mm MAPI single crystal at 296K and 74K using a Rigaku RU-300B x-ray source coupled to a eulerian 4-Circle Huber diffractometer. A focusing multilayer monochromator was used to select the K$\alpha$ emission line of the Cu anticathode ($\lambda$ =1.54 Å), and generate a 0.8 mm beam size at sample position. The sample was cooled down with a CryoVac open-circuit $LN_2$ cryostat. $\theta$-$2\theta$ scans were measured in specular geometry, confirming the [100] orientation of the sample surface.

**Optical spectroscopy:** Absorption was measured inside a Janis cryostat inserted in a Perkin-Elmer spectrophotometer. The photoluminescence spectra were recorded using a Spectrapro 2500i spectrometer equipped with a Pixis: 100B CCD array detector (Ropers Scientific). The excitation wavelength was the 325 nm (3.815 eV) line of a He-Cd laser. The sample was placed in a closed cycle cryostat.

Time-resolved photoluminescence were performed using the Time-correlated single Photon counting TimeHarp 260 system from PicoQuant. The excitation was the second harmonic of a pulse from a Ti:Sapphire laser (Mai Tai, Spectra-Physics). The emission was detected with a single photon avalanche diode (IDQuantique ID150).

## 3. Results and discussion

Millimeter-sized single crystals and thin films of $CH_3NH_3PbI_3$ has been prepared using procedure described in section 2. Figure 1 shows SEM views of the morphology of a single crystal and a thin polycrystalline film. Despite some local inhomogeneities, single crystals present clean and relatively flat surfaces with typical length superior to the excitation spot diameter used for PL ($d \approx 40\ \mu m$) (Figure 1(b)). Hence, we could probe the emission properties of a large monocrystalline domain of $CH_3NH_3PbI_3$ and compare the results with the luminescence of thin films presenting relatively small grains.

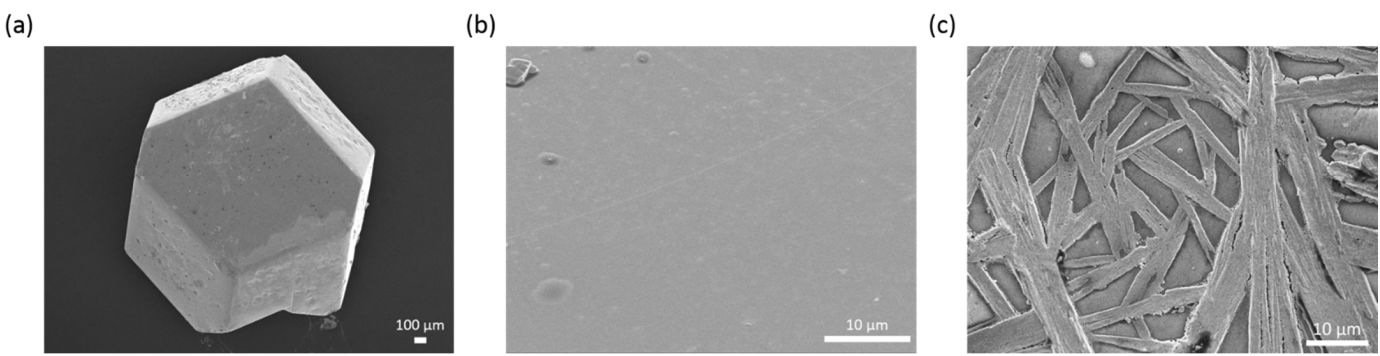

**Figure 1**: (a) Top view SEM image of a mm-sized $CH_3NH_3PbI_3$ crystal (b) Higher resolution SEM image of the surface of the crystal. (c) SEM view of the surface of a $CH_3NH_3PbI_3$ thin film.

Figure 2 presents the PL spectra of a single crystal of $CH_3NH_3PbI_3$, in red, between 10K and 290K measured under weak continuous irradiation (8 W/cm$^2$). Spectra of a thin film of $CH_3NH_3PbI_3$ is displayed for comparison in black. Between 10K and 80K, the PL emission of crystals can be divided in a sharp line at high energy, which we will be tentatively assigned to Free Excitonic (FE) emission and a multi-components broad emission at lower energy (LE), related to trap states. The LE emission quenches at higher temperature and is very weak at 80K. Between 10K and 160K, the FE emission presents an important broadening with the temperature. At approximately 140K, a new contribution starts to grow, positioned at approximately 1.58 eV, which correspond to the emission of the room temperature tetragonal phase. This emission becomes dominant above 160K, while the FE band disappears. The phenomena is due to the tetragonal-orthorhombic phase transition which induces a band gap energy switching of 100 meV.[9,17] A single broad emission band (T Band) is seeable, above 180K, in the room temperature tetragonal phase. The phase transition has interesting and complex effects on the optical properties of $CH_3NH_3PbI_3$ and will be discussed in details elsewhere.

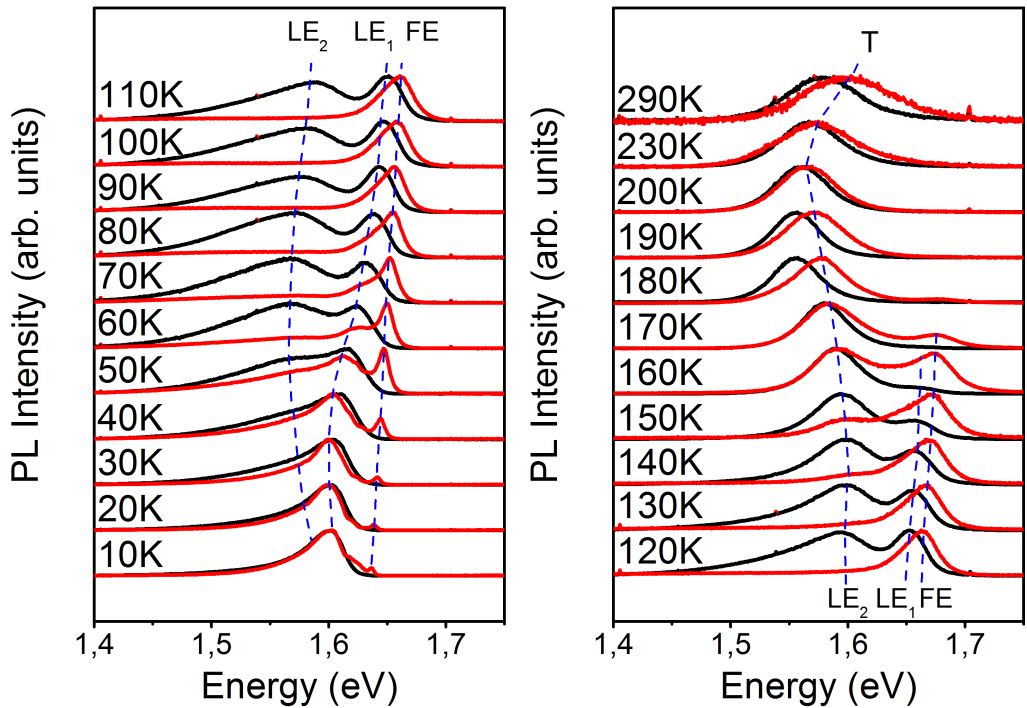

**Figure 2**: PL spectra at different temperature of $CH_3NH_3PbI_3$ in the form of thin film (black) and crystal (red), under weak excitation of a continuous wave laser at 405nm (8 W/cm$^2$). Dashed lines are visual indications of the evolution of the main PL bands (see text).

If we turn to the thin film emission, we observe that the FE band is completely absent on the spectra, at any temperature. Thin films present only a broad multi-component band at low energy, which share similar characteristics (energy position and width) with the LE emission of the $CH_3NH_3PbI_3$ crystal. In particular, at 10K, the LE emission band, positioned at 1.60 eV and with a FWHM of ca. 60 meV superimposes very well with the LE emission of the crystal. Above 30K, two bands emerges from LE emission on both films and crystals (hereafter denoted $LE_1$ and $LE_2$, see Figure 2). We note that the $LE_2$ emission is quite stronger on thin films and seems to extend in the high temperature tetragonal phase. This last observation is in good agreement with previous measurements on thin films and has led to the hypothesis of inclusions of the tetragonal phase in the low temperature orthorhombic phase, due to an incomplete phase transition.[8–10] The $LE_2$ emission, though weak and almost undistinguishable above 60K, is still present on crystals spectra, indicating that coexistence of phase may exist for the crystalline form. We have performed measurements at different points of the crystal and have observed for specific positions a strong $LE_2$ component. Note that the as-grown crystals present surface imperfections (Figure 1(a)). Our measures suggest that small tetragonal phase domains exist at low temperature at the surface of crystals, possibly due to local surface imperfections. Crystal imperfections might help to stabilize the high temperature phase through mechanical strain. In fact, the phase transition in $CH_3NH_3PbI_3$ has a mechanical origin and different phases could be stabilized by strain.[31] The coexistence of different phases has also been demonstrated in inorganic perovskites.[32–34]

To further analyze the nature of the PL emission, we measure spectra at different power densities of a femtosecond laser for $CH_3NH_3PbI_3$ crystals and thin films at 10K, 50K and 100K (Figure 3). On the crystal spectra, the sharp FE emission clearly dominates the spectra at high fluence on crystals and maintains its position. The $LE_2$ band presents a relative decrease at higher laser density, compatible with a saturation behavior, and shifts toward higher energy. On the contrary, the $LE_1$ band seems to maintain its position. At 50K, this component is clearly visible at 1.619 eV and at 100K, the $LE_1$ emission appears as a shoulder on the left of FE peak, at ca. 1.65 eV.

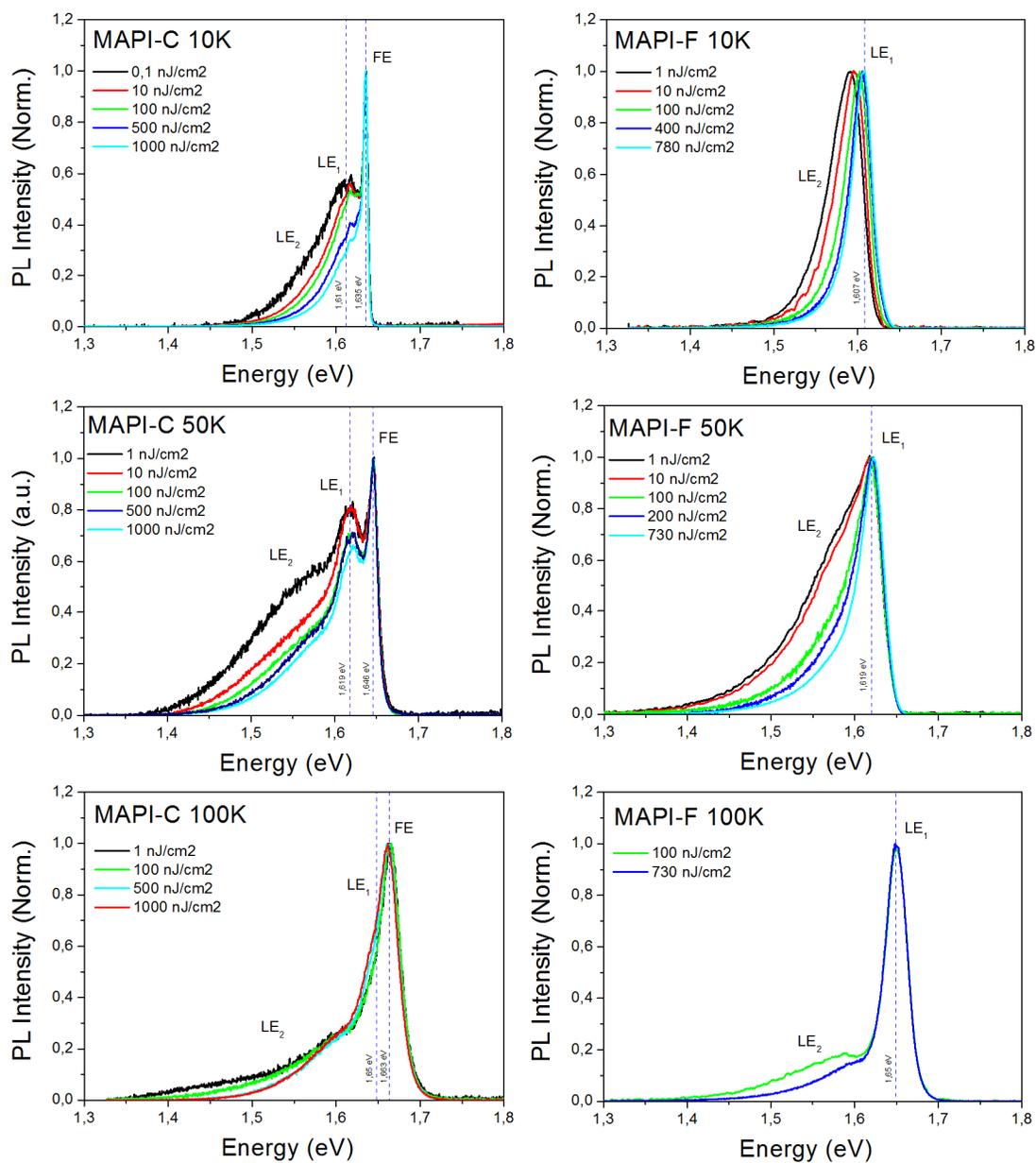

**Figure 3**: Power dependence of the PL of $CH_3NH_3PbI_3$ single crystal (MAPI-C, left) and thin film (MAPI-F, right) at different temperatures.

On thin films, the high energy FE peak is completely absent, even at the highest fluence. The film spectra show only the $LE_1$ and $LE_2$ bands, with similar energy position, width, and fluence-dependence than the previously described LE bands of the crystal. We remark that the PL spectra and fluence-dependence of thin films is consistent with previous observations.[35]

Figure 4 shows the power density dependence of the PL Integrated Intensity of the FE emission at 10K. The intensity could be fitted by a power law of the form $I^\lambda$ with $\lambda = 1.17$. In direct band gap semiconductor, $\lambda$ should be comprised between 1 and 2 for excitonic recombination.[36] This result confirms the excitonic nature of the emission.

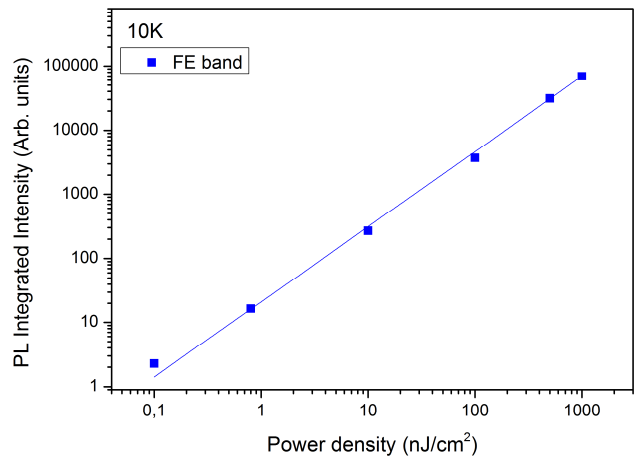

**Figure 4**: PL Integrated Intensity as function of laser power density of the free excitonic emission on single crystal. The line is a linear fit with slope equal to 1.17.

We focus now on the thermal evolution of the FE line between 10K and 160K. The PL spectra has been consistently decomposed using multi-peak fitting, in order to extract the PL integrated intensity and Full-Width-at-Half-Maximum (FWHM) of the FE emission (See details of the procedure in SI and deconvolution examples in Figure S2)

Figure 5(a) shows the evolution of the FWHM of the FE emission as a function of temperature. As previously stated, the FE line presents a FWHM of only 5 meV at 10K. The observation of a narrow linewidth at low temperature is an indication of a weak inhomogeneous broadening and of the good crystallinity of our sample. Such narrow linewidth could be compared to the typical linewidth of excitonic peaks in high quality bulk inorganic semiconductor, which is approximately 1 meV at temperature close to 10K (see for example in GaN[37,38], ZnO,[39,40] and GaAs[41]). We note that narrow free excitonic emissions of a ca. 5 meV have also been reported in $CsPbX_3$ (X=Cl,Br)[42,43] and $PbI_2$.[44]

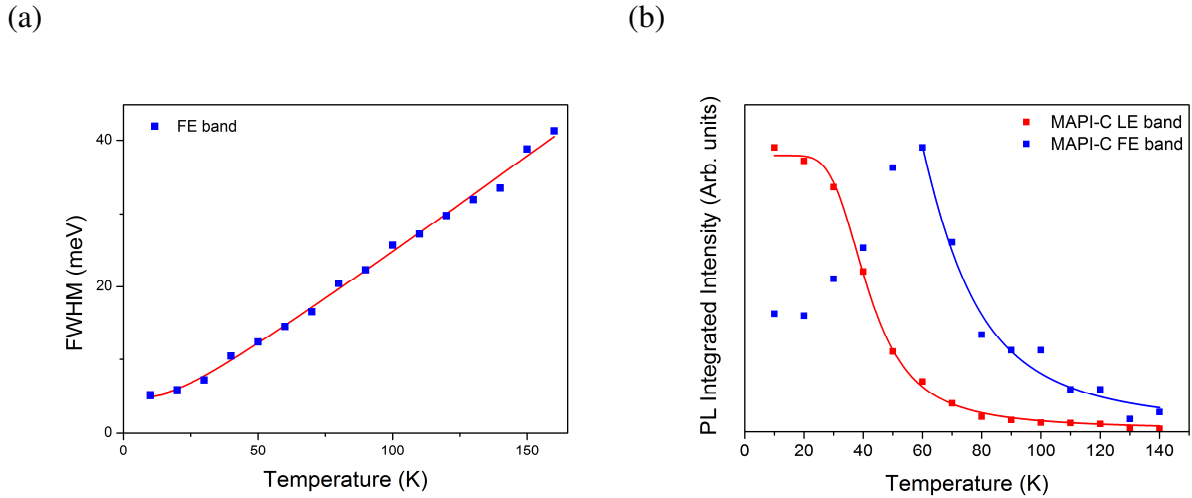

**Figure 5**: (a) Full Width at Half Maximum (FWHM) of the Free Excitonic (FE) emission of $CH_3NH_3PbI_3$ crystal as function of temperature. The fit of the FE emission, with equation (1) is indicated in red. (b) PL Integrated Intensity of the FE emission (blue) and of the Low Energy (LE) emissions of $CH_3NH_3PbI_3$ crystal (MAPI-C) as function of temperature. Lines are fit of the data using equation (2) (see text.)

We measure a large broadening of the FE emission with temperature, from 5 meV at 10K to 40 meV at 160K. This broadening is quite stronger than in typical semiconductors in the same thermal range. For example, in the inorganic direct band gap semiconductor GaAs, between 10K and 150K, the linewidth increases very slightly of approximately a factor 2 (from approximately 1.5 to 3 meV).[45] Hence, the thermal evolution of the FE linewidth suggest a strong homogenous broadening due electron-phonon coupling. Note that a strong phonon coupling in $CH_3NH_3PbI_3$ has been hypothesized to explain the large homogenous broadening observed at room temperature.[46]

The following relation has been proposed to describe the evolution of exciton linewidth with temperature in semiconductors:[47]

$$\Gamma(T) = \Gamma_0 + \gamma_{ac}T + \gamma_{LO}/\left(e^{E_{LO}/kT} - 1\right) + \gamma_{imp}e^{-E_b/kT} \qquad (1)$$

$\Gamma_0$ is the temperature independent inhomogenous broadening, the terms $\gamma_{ac}, \gamma_{LO}$ arise from the scattering by the acoustic phonons and the optical phonons respectively. $E_{LO}$ is the energy of the optical phonons. The last term account for the inhomogeneous broadening due to scattering with ionized impurities. Phonon energies are still not well established in 3D HOP[48,49] and we are forced to make assumptions. First, we remark that the impurities term failed to reproduce the linear behavior observed above 20K (Magenta dashed line Figure S3). The result suggests that scattering with impurities is not the dominant broadening mechanism, in agreement with conclusions based on analysis of the thermal dependence of charge-carriers mobility[50] and PL linewidth in thin films.[51] We will neglect the term in the following analysis. Acoustic phonons are generally the dominant term in typical semiconductors below 100K, and optical phonons mechanism is generally negligible.[47] Due to the relative linearity of the excitonic FWHM curve, the data could be fitted approximately with only the two first terms of Eq.1 (Green dashed line in Figure S3). It results in a value of $\gamma_{ac} = 245 \pm 5\ \mu eV.K^{-1}$, more than an order of magnitude higher than in bulk GaAs.[52] In 2D HOP, which self-assemble in natural quantum well structures, an acoustic phonon coupling ten times greater than in GaAs quantum wells has been reported.[53] However, the quality of the fit is not satisfying at 10K.

On the other hand, if we suppose that the optical phonons are active in this low temperature range, the acoustic phonon term becomes negligible and the equation (1) could reproduce the data with a better result, especially at 10K, with the following values $\Gamma_0 = 4.9 \pm 0.3$ meV, $\gamma_{LO} = 14.4 \pm$ $0.3\ meV$ and $E_{LO} = 4.7 \pm 0.9\ meV$ (solid red line on Figure 5(a)). The latter energy is weak compared to the typical values obtained for inorganic semiconductors.[47] Recent studies based on experiments on thin films and theoretical considerations estimated the LO phonon energy to approximately 10 meV.[49,51] Our result is somewhat smaller but is in very good agreement with a recent estimation of an effective polaronic LO phonon energy of 4.1 meV by Soufiani *et al.*[21] deduced from permittivity measurements.

We discuss of the thermal evolution of the PL intensity. Figure 5(b) represents the PL integrated intensity of the free excitonic emission (blue curve) and of the low energy band of CH$_3$NH$_3$PbI$_3$ single crystals. The PL intensity of the FE peak presents a non-monotonous thermal evolution: it increases first with temperature and reaches a maximum at 50K and then decreases. This behavior is associated with the quenching of the low energy defect emissions (red dot). The thermal quenching of the PL integrated intensity could be described with the familiar equation:

$$I_{PL} = \frac{I_0}{1+Ae^{-\frac{E_a}{k_B T}}} \quad (2)$$

Where $I_0$ is the PL intensity at 0K. $E_a$ is the activation energy of a non-radiative process and $k_B$ is the Boltzmann constant.

We estimate an activation energy of approximately $E_a=19.9\pm 0.6$ meV for the quenching of the low energy band, associated to the exciton detrapping (solid red line on Figure 5(b)). Above 50K, the FE emission quenches in turn with an activation energy of $E_a=28.4 \pm 0.6$ meV. The latter process could be connected to the dissociation of the exciton and the energy corresponds to the binding energy of the free exciton in the low temperature orthorhombic phase. Our estimation of the binding energy in the low temperature phase is in good agreement with recent studies, based on the fitting of the thin films absorption spectrum with Elliott's formulas : Yamada *et al.*[24] reports an energy of 30 meV, Saba *et al.* and Soufiani *et al.* estimate the binding energy at approximately 25 meV.[21,54] However, Nicholas and coworkers use a more direct method to estimate the binding energy, relying on the analysis of thin films absorption spectra under high magnetic field and report a binding energy of only 16 meV in the orthorhombic phase.[55] We note that, to compare those values, we must ensure that we are really looking at the same material. As previously mentioned, the optical and electronic properties of MAPI present important variation with synthesis conditions. Petrozza and coworkers have shown that the intensity of the excitonic peak on the absorption spectra of thin films, measured at 4K, depends on the degree of crystallinity of the material and could eventually vanished for low crystallinity films.[7] They suggest that the degree of order-disorder in the films, especially the order of the organic cation, may have a large impact on exciton binding energy.[14]

We performed Time-Resolved Photoluminescence (TRPL) measurements to get more insight on the recombination mechanisms. Figure 6(a) presents the PL dynamics decay at 10K of the FE emission on MAPI-C and at the maximum of the LE band on crystals and films. The decay of the FE presents a very fast, non-exponential dynamics. The PL dynamic could be fitted with the convolution of the Instrumental Response Function (IRF) with four exponentials. The results are presented in Table 1. The PL lifetime of the FE emission presents a sub-nanosecond dynamics with a leading time shorter than our time resolution (75 ps). For comparison, fast decay with leading time of the order of several tens of ps has been reported for the free excitonic emission of single crystals of $CsPbX_3(X=Cl,Br)$.[43,56] A non-exponential decay with an approximately 100 ps characteristic time has also been reported for single crystal of $PbI_2$.[57] This fast, non-exponential recombination may result from efficient capture processes by trap states. In fact, we observed a rising time on the LE emission dynamics (Figure 6(b)). The trap states on both crystals and films present at 10K a very similar decay with ns timescale and could be fitted with three exponentials. We note that the estimation of the slower component is limited by the time range of our measurement. Fang *et al.* report a millisecond component for the low energy emission in crystal, which they attribute to triplet state.[17]

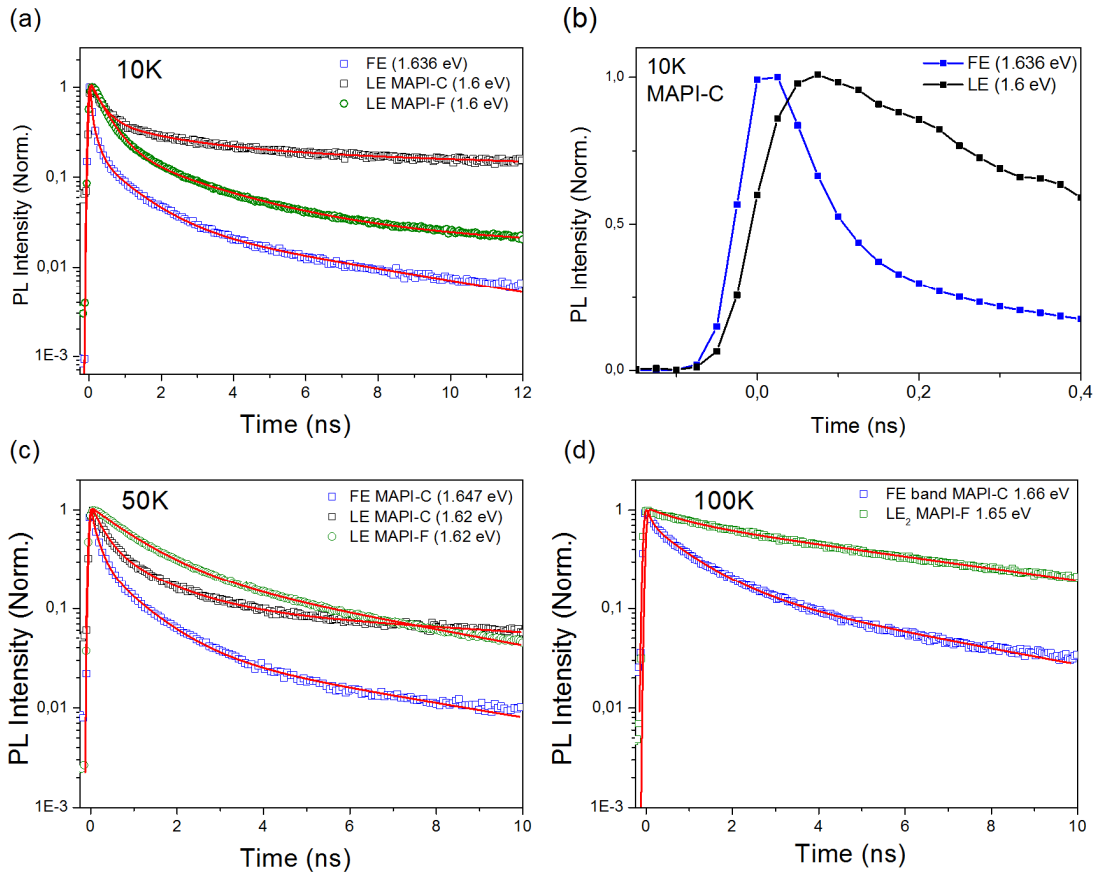

**Figure 6**: Time-Resolved Photoluminescence at different temperatures. (a) Comparison of the PL decay at 10K of the FE emission and of the LE emission (at 1.6 eV) of $CH_3NH_3PbI_3$ film (MAPI-F) and crystal (MAPI-C). (b) Comparison of the rising time at 10K of the FE and LE emission of $CH_3NH_3PbI_3$ crystal (c) FE and LE band decay on crystal and film. (d) Decay at 100K of the FE emission of crystal and $LE_1$ band of thin film. Fitting curves are indicated by red lines. Results of the fit are indicated in Table 1

The PL lifetime of the emissions increases with the temperature (Figure 6(c) and 6(d)) and the PL decay is less multi-exponential: at 100K, the LE band decay is slightly bi-exponential. In particular, the picosecond component of FE decay tends to disappear above 50K, concomitantly with the previously described quenching of the trap states emission. Thus, the thermal evolution of the recombination dynamics supports the idea that the fast components of the FE emission are dominated by the trapping time of the excitons. Above 50K, the capture of excitons by trap states becomes less efficient and it results in a slower dynamics. Thus, the slower component of the PL decay is linked to the recombination time of the trap states. A fast, sub-nanosecond component in the PL of $CH_3NH_3PbI_3$ thin films, has been recently reported at room temperature, under weak irradiation, attributed to a rapid relaxation of the carriers to defects states.[58]

**Table 1**: Results of the TRPL fits on $CH_3NH_3PbI_3$ single crystal (MAPI-C) and thin film (MAPI-F) at different temperatures.

|  |  | Emission Energy (eV) | $t_1$ (ns) | $t_2$ (ns) | $t_3$ (ns) | $t_4$ (ns) |
|---|---|---|---|---|---|---|
| 10 K | MAPI-C | 1.66 eV (FE band) | < 0.075 | 0.15 | 1.01 | 6.59 |
|  |  | 1.6 eV (LE band) | 0.33 | 2.15 | 38.86 | - |
|  | MAPI-F | 1.6 eV (LE band) | 0.44 | 2.56 | 35 | - |
| 50 K | MAPI-C | 1.65 eV (FE band) | < 0.075 | 0.17 | 0.91 | 6.02 |
|  |  | 1.62 eV (LE band) | 0.26 | 1.25 | 16.76 | - |
|  | MAPI-F | 1.62 eV (LE band) | 1.09 | 5.5 | - | - |
| 100 K | MAPI-C | 1.66 eV (FE band) | 0.12 | 1.02 | 5.44 | - |
|  | MAPI-F | 1.65 eV (LE band) | 0.90 | 7.17 | - | - |

We will, in the next paragraph, discuss our results taken together. From our findings, two questions naturally arise: what could explain (1) the absence of FE emission in the PL of polycrystalline thin films (2) the strong similarities of the trap states emission on films and crystals. To answer those questions, we may consider several hypothesis.

For the first point, based on the studies of A. Petrozza and coworkers, we may suppose that the crystallinity of our thin film is insufficient to stabilize excitons, even at low temperature. However, this possibility could be ruled out, as low temperature absorption spectra clearly indicate an excitonic resonance (Figure S4) on our sample. A simplest approach would be to consider that the excitons trapping is far more efficient in polycrystalline thin films, compared to single crystals, and is responsible for the quenching of the FE emission. This hypothesis led us to the question of the nature of those trap states, and therefore to the second point of our discussion.

Excitons could be trapped by defects, impurities or structural disorder in general. We emphasize again the strong difference between the microstructure of the thin films and crystals, illustrated in Figure 1. A greater number of trap states are likely to form in disordered, polycrystalline thin films. However, the striking correspondence between the LE emission of thin films and crystals at low temperature seems to imply the existence of a finite number of well-defined trap states, with a same physical origin on both compounds, rather than a collection of trap states with variable depth.

As previously discussed, the $LE_2$ band is mainly due to the presence of small tetragonal phase domains. However, those inclusions could not account for the whole LE emission, in particular for the $LE_1$ band. A possible origin for the $LE_1$ band may arise from polaronic effect. In the polaron framework, strong coupling of an exciton/carrier with the lattice vibration could lead to its immobilization (ie self-trapping). Self-trapping induces broad, Stokes-shifted emissions and occurs in pico or sub-picosecond timescale.[59] The phenomena has been observed in alkali-halide and lead-halide[59,60] and may account for white-light emission in 2D HOP.[61–63] Polaronic effects have been invoked in different studies to explain the optical and transport properties of 3D HOP.[21,46,64–66] Many authors suggest that the reorientation of the MA cation and its associated dipole, is connected to the formation of these polaronic states.[64,67,68] The strong broadening, associated to LO phonons coupling, observed in this study and the absence of saturation at high fluence is consistent with the formation of self-trapped excitons/carriers. Importantly, the self-trapping could be intrinsic or extrinsic. In the latter case, the process is mediated by the presence of disorder and/or defects.[69] Hence, the self-trapping of excitons/carries may likely be formed at the surface or interfaces between grains, where reorientation of the terminal MA cations and deformation of the crystal structure in general is easier, as also proposed by Wu *et al.*[65]

In thin polycrystalline films, with relatively small micrometric grain size, we suppose a fast diffusion and self-trapping of excitons at the interfaces, which completely prevent the observation of a free excitonic emission even at high laser power density. The surface/interface origin of the trap states could explain the important similarity between the LE emission band of films and crystals: due to the large absorption coefficient of $CH_3NH_3PbI_3$ ($\alpha \approx 10^5 cm^{-1}$ for an excitation wavelength of 400nm)[70], the penetration depth of the light is approximately of 100nm, which means that we are still sensitive to the surface states, despite the large size of the crystal. As previously mentioned, crystals present locally some inhomogeneities and roughness and we have observed relative variation of the free excitonic emission compared to the LE band intensity at some points of the crystal surface. However, crystals, unlike thin films, present also large areas with clean and smooth surfaces giving access to the intrinsic emission properties of $CH_3NH_3PbI_3$.

## Conclusions

$CH_3NH_3PbI_3$ single crystal emission spectra are composed, at low temperature, of a sharp line, with a FWHM of 5 meV at 10K, assigned to Free Excitonic emission and broad band emission at low energy. From the analysis of the thermal evolution of the free excitonic line, we demonstrate a strong homogenous broadening of the emission connected to LO phonon interaction and extract a binding energy of approximately 28 meV. The FE emission presents a non-exponential decay with sub-nanosecond leading time, dominated by the rapid relaxation to trap states. In comparison, the emission of MAPI thin films at low temperature is dominated by trap states and the free excitonic emission is never seen even at high fluence. The low energy, trap states band are very similar at low temperature on thin films and crystal in term of position, width and recombination dynamics and is likely connected to formation of trap states at the surface/interfaces of the material. Our results lead to a deeper understanding of the photophysics of 3D HOP and highlight the importance of studying bulk single crystals in order to access to the intrinsic properties of 3D HOP. Besides, our findings reinforce the idea that low temperature PL spectroscopy is a relevant tool to assess the crystalline quality in 3D HOP.

## Acknowledgments

This work has received funding from the European Union's Horizon 2020 research and innovation programme under the grant agreement N° 687008. The information and views set out in this paper are those of the author(s) and do not necessarily reflect the official opinion of the European Union. Neither the European Union institutions and bodies nor any person acting on their behalf may be held responsible for the use which may be made of the information contained herein.

# Supplementary Information

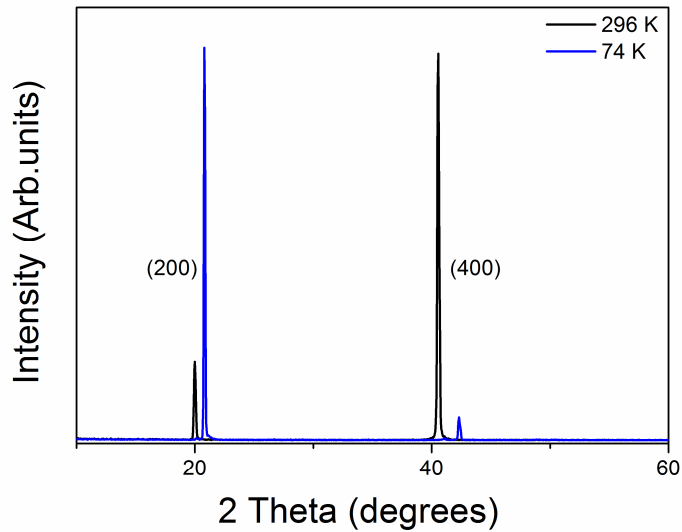

**Figure S1**: XRD pattern obtained from a single crystal oriented along the [200] direction at room temperature and at 74K.

**X-ray diffraction**: The XRD patterns measured in the tetragonal and orthorhombic phases (296K and 74K respectively) present sharp peaks limited by instrumental resolution, with positions in good agreement with previous reports of $CH_3NH_3PbI_3$ structure at those temperatures (Figure S1).[1,2]

**Deconvolution procedure.** Exciton line shape have generally both homogeneous and inhomogeneous parts. In typical semiconductors, the inhomogeneous part is related to disorder and impurities while the homogenous part is mainly determined by exciton scattering with acoustic and optical phonons. Inhomogeneous broadening is associated with a Gaussian shape, while homogenous broadening induce a Lorentzian shape.[3] Hence, the photoluminescence spectra has been consistently decomposed using Voigt functions. To improve the precision of our adjustment, multiple data set have been fitted, with different fluences. Global fitting of spectra with different fluences allow to improve the precision. We also have measured the photoluminescence as function of temperature on several samples and realized the fitting procedure on the different datasets.

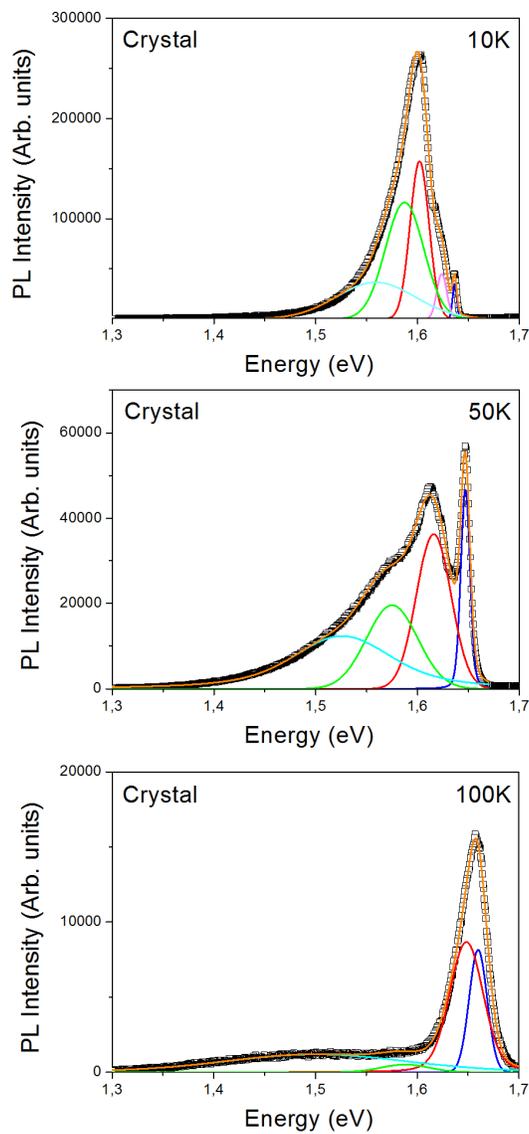

**Figure S2**: Fit of the PL spectra with Voigt functions of $CH_3NH_3PbI_3$ crystal at different temperatures

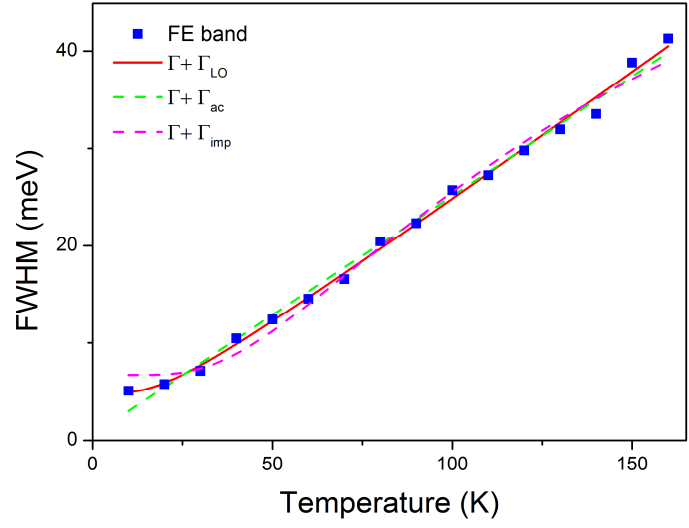

**Figure S3**: FWHM of the Free Excitonic emission on $CH_3NH_3PbI_3$ single crystals as function of temperature fitted with the different terms of equation (1) (Green dashed line, $\Gamma_0 = 0.5 \pm 0.5 \ meV$, $\gamma_{ac} = 245 \pm 5 \ \mu eV.K^{-1}$; Magenta dashed line: $\Gamma_0 = 6.7 \pm 0.5 \ meV$, $\gamma_{imp} = 79.2 \pm 0.4 \ \mu eV.K^{-1}$, $E_b = 12.3 \pm 0.1 \ meV$ ; solid line, $\Gamma_0 = 4.9 \pm 0.3 \ meV$, $\gamma_{ac} = 0 \ meV$, $\gamma_{LO} = 14.4 \pm 0.3 \ meV$, $E_{LO} = 4.7 \pm 0.9 \ meV$).

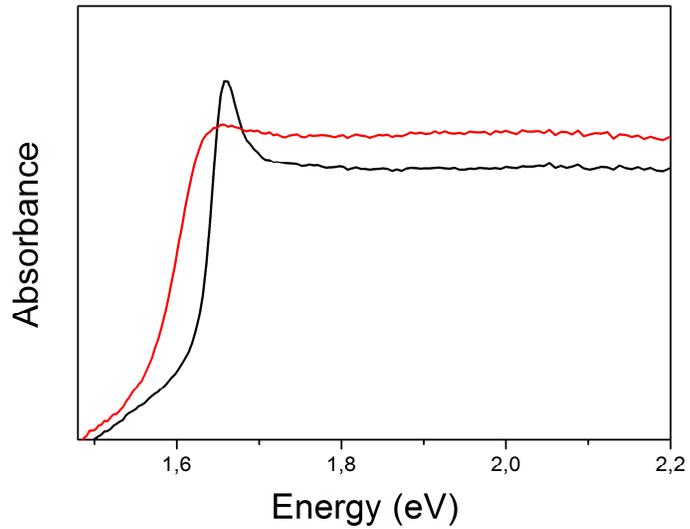

**Figure S4**: Optical absorption spectra of $CH_3NH_3PbI_3$ thin film at 290K (red) and 50K (black)